\documentclass[12pt]{article}
\usepackage{amsgen}
\usepackage{amsfonts}
\usepackage{amssymb}
\usepackage{amsbsy}

\begin{document}

\title{On group averaging for non--compact groups}

\author{Andr\'es Gomberoff\footnote{email: {\tt andres@cecs.cl}}\\
Centro de Estudios Cient\'{\i}ficos, Casilla 1469, Valdivia, Chile.\\ Physics
Department, Syracuse University, Syracuse,   New York 13244, USA.}

\maketitle

\begin{abstract}
We review some aspects of the use of a  technique known as group averaging,
which  provides a tool for the study of constrained systems.  We focus our attention
on the case where the gauge group is non--compact, and  a `renormalized' group averaging method
must be introduced.  We discuss  the connection between superselection
sectors and the rate of divergence of the group averaging integral.
\end{abstract}
\section{Introduction}

What follows is primarily based in the article \cite{ggmm} written in collaboration
with Donald Marolf.  A short review of some  previous work  has been added
in order to make it self--contained. The focus has also been changed, stressing the physical
issues, and referring the reader to the original literature for details on the mathematical
technology.

Quantization of  constrained systems was first studied by Dirac\cite{Dirac}(a complete
treatment of the subject can be found in Ref. \cite{HT}).
His method involves introducing the constraints as operators on some space.
 The {\em physical Hilbert space} is then defined to contain only those states
annihilated by all the constraints. This last step is the one in which we will concentrate our attention.
In particular, we will discuss the conditions under which a {\em physical inner product} can be
constructed  explicitly.
There are many ways of  implementing the Dirac method. Here we are interested in
what is called {\em refined algebraic quantization} (RAQ) \cite{ALMMT,GM,KL,AH,QORD},
which is particularly suitable for dealing with canonical quantization within a generally applicable
mathematical framework.

RAQ becomes much more powerful when
a technique known as {\em group averaging}
can be applied.  Group averaging uses
the integral
\begin{equation}
\label{GAI}
I=\int_G \langle \phi_1|U(g)|\phi_2 \rangle \ dg
\end{equation}
over the gauge group $G$
to define the physical Hilbert space.  Here  $dg$ is a
Haar measure on $G$. Once a space of states
 has been found for which this procedure converges, group
averaging gives an {\it algorithm} for the implementation of RAQ.  When
group averaging converges sufficiently strongly this
algorithm gives the {\it unique} implementation of RAQ \cite{GM2}.
However, it will often happen that group averaging fails to converge
on some interesting domain.   As described in \cite{GM2},
the fact that convergent group averaging ensures a unique
representation (compatible with RAQ) of the algebra of observables
shows that group averaging {\it must} in
fact diverge in the presence of any superselection rules.  However, as
was described in \cite{ALMMT}, one can sometimes construct a {\em renormalized}
group averaging operation, even when group averaging
does not properly converge.
Our goal here is to study the possibility of implementing a general method for
constructing renormalized group averaging  integrals. We will  discuss  in parallel
the particular case fully treated in \cite{ggmm}, where the method has been successfully
applied to the $SO(n,1)$ group acting on Minkowski space.

We will start in section \ref{RAQ} with a short and simplified review of RAQ. In section \ref{convergency} we will
study the convergency properties of the integrals of the form (\ref{GAI}). In section \ref{regularization} we will discuss the
regularization of the divergent group averaging integrals. In section \ref{superselected} we will show the emergence of
superselected sectors, and we will finish  in section \ref{conc} with conclusions and final comments.

\section{Refined Algebraic Quantization}
\label{RAQ}

The starting point of Refined Algebraic Quantization is an  {\em auxiliary}  Hilbert space
${\cal H}_{\mbox{\tiny aux}}$ with an inner product $\langle,\rangle$. We also have a set of  constraints
$C^{A}=0$ generating a Lie Group $G$, a unitary representation of it, $U(g)$,  acting on
${\cal H}_{\mbox{\tiny aux}}$ and an algebra  of selfadjoint observables ${\cal A}$ commuting
with all the constraints.
The core of the method stands in finding the so called {\em rigging map} $\eta$,
\begin{equation}
\begin{array}{cccc}
\eta :& \ {\cal H}_{\mbox{\tiny aux}} & \longrightarrow & ^*{\cal H}_{\mbox{\tiny aux}} \\
         &|\phi\rangle & \longmapsto & _{\mbox{\tiny phy}}\!\langle\phi|
\end{array}
\end{equation}
such that
\begin{equation}
_{\mbox{\tiny phy}}\!\langle\phi| C^{A} =0 \ .
\label{rigging}
\end{equation}

Here $^{*}{ \cal H}_{\mbox{\tiny aux}}$ is the dual of ${ \cal H}_{\mbox{\tiny aux}}$\footnote{This is an
  simplification of the actual definition, but is enough for our purpose here.  Rigorously speaking, the rigging map goes
 from a dense subspace $\Phi$ of   ${ \cal H}_{\mbox{\tiny aux}}$ to its dual $^{*}\Phi$
 (see for example Ref. \cite{ALMMT}). }. The rigging  map $\eta$ defines both  the {\em physical} Hilbert space
 ${\cal H_{\mbox{\tiny phy}}}$
and its inner product. ${\cal H_{\mbox{\tiny phy}}}$ is simply the image of the rigging map, and
\begin{equation}
\langle\phi_1|\phi_2\rangle_{\mbox{\tiny phy}} = \eta(|\phi_1\rangle)|\phi_2\rangle \ \ .
\label{inner}
\end{equation}
The rigging map must satisfy two requirements: It must be such that (\ref{inner}) is a definite positive
inner product and must commute with all the observables.

It has been shown\cite{GM2} that if  the integral $I$ in (\ref{GAI})
converges for all $|\phi_1\rangle$ and $|\phi_2\rangle$  then the rigging map is uniquely determined.
The inner product (\ref{inner}) is then given precisely by the integral (\ref{GAI}). This
procedure for implementing the rigging map is known as {\em group averaging}.

\section{On convergence of the group averaging integrals}
\label{convergency}

Let us now focus our attention in the convergence of the  integral $I$ in (\ref{GAI}).
\begin{equation}
I \le \int_G \left|\langle \phi_1|U(g)|\phi_2 \rangle\right| \ dg = \int_G dg = V_{G} \ \ ,
\label{conv}
\end{equation}
and therefore the integral is bounded by the volume of $G$ and it is guaranteed  to converge for any compact
group. The integral will diverge, in the worst case, as the volume of the group. This will happen for example if  we plug
in Eq. (\ref{GAI})  $\ |\phi_2\rangle=|\phi_1\rangle=|\psi\rangle$ for some $G$--invariant state $|\psi\rangle$.
For non--compact groups there are some particular cases where the group averaging procedure converges (see for example
\cite{BC}), but  generically, convergency will fail. Our goal is to find some regularization procedure in order to define a finite,
(renormalized)  generalization of $I$.

\section{Regularization of the group averaging integrals}
\label{regularization}

The possibility of regularizing the integral $I$  lays in our ability to find a parameter, $\lambda\in [0,\infty]$, such that:
\begin{itemize}
\item $G(\lambda)$ is a finite domain on $G$ such that $G(\lambda) \rightarrow G$
when $\lambda\rightarrow \infty$.
\item $\langle\phi_1|\phi_2\rangle_{\lambda}=\int_{G(\lambda)} \langle \phi_1|U(g)|\phi_2 \rangle \ dg \longrightarrow
f^i(\phi_1,\phi_2)\Lambda^{i}(\lambda)$ \\ for big values of $\lambda$,
\end{itemize}
where $f^i(\phi_1,\phi_2)$ are finite functions of the states and $\Lambda^i(\lambda)$ diverges for
$\lambda\rightarrow\infty$. $\Lambda^i (\lambda)$  may depend on the states only through the integer $i$, which labels
different degrees of divergency that might occur.

If we can find such parameter we will define:
\begin{equation}
\langle\phi_1|\phi_2\rangle_{\mbox{\tiny phy}}=f^i(\phi_1,\phi_2)  \ .
\label{newrigging}
\end{equation}

We do not know a general method for obtaining $G(\lambda)$. A particular example has been given
in \cite{ggmm} for the Lorentz group $SO(n,1)$. In that case, however, the function $G(\lambda)$ was also a function of
the states $|\phi_i\rangle$. This adds more difficulty to our problem, because one has to check explicitely the consistency of the inner product with the  requirements of Refined Algebraic Quantization which are automatically satisfied in the present case.

\section{The normalized group averaging integral and superselected sectors}
\label{superselected}

Let us now study the consequences of the above definition of the rigging map.
In general, $U(g)$ does not have to be an irreducible representation. For instance, if we take $SO(n,1)$ acting on
an auxiliary Hilbert space defined by the set of functions with compact support on Minkowski space ${\bf
R}^{2+1}$,  any hyperboloid $x_1^2+\cdots x_n^2-t^2=$constant is an invariant subspace where the action of the group is
irreducible.
The auxiliary Hilbert space can be written as a direct sum:
\begin{equation}
{\cal H}_{\mbox{\tiny aux}}= \bigoplus_{a}{\cal H}_a
\label{hs}
\end{equation}

Now we will show that for two states $|\chi_1\rangle$ and $|\chi_2\rangle$ belonging to the same irreducible subspace,
the degree of divergency of $\langle\chi_1|\chi_1\rangle_\lambda$ and  $\langle\chi_2|\chi_2\rangle_\lambda$ is the same. In
fact, as there
exist some $g_0$ such that $|\chi_2\rangle=U(g_0)|\chi_1\rangle$,
$$
 \langle\chi_2|\chi_2\rangle_\lambda = \int_{G_\lambda} \langle \chi_1|U(g_0)^{\dagger}U(g)U(g_0)|\chi_1 \rangle \ dg \ ,
$$
and because we can go continuously from the identity to $g_0$ for any $\lambda$, the function $\Lambda(\lambda)$
cannot  have a discontinuity from some $\Lambda^i(\lambda)$ to another function $\Lambda^j(\lambda)$ with different
asymptotic behavior. By a similar argument it is straightforward to note that given any two states in some ${\cal H}_a$,
their inner product $\langle\chi_i|\chi_j\rangle_\lambda $ is either zero or shows the same degree of divergency.
Now we  rewrite the auxiliary  Hilbert space as a direct sum
\begin{equation}
{\cal H}_{\mbox{\tiny aux}}= \bigoplus_{i}{\cal H}_i  \ ,
\label{hs2}
\end{equation}
where we have just grouped together the ${\cal H}_{a}$ showing the same degree of divergency in $\lambda$.
For the $SO(n,1)$ case, the Hilbert space is divided in two sectors, defined by functions supported inside or outside
the light cone\cite{ggmm}. The group average integrals converge inside the light cone. Outside, the integrals diverge as $\lambda^{n-2}$ for $n>2$ and as $\log\lambda$ for $n=2$. For $\lambda=1$ they converge.
The physical inner product is given by
$$
\langle x|y\rangle_{\mbox{\tiny phy}} =\frac{1}{(x^2)^{(n-1)/2}}\delta(x^2-y^2) \ \ ,
$$
where $x^2 = x_1^2+\cdots x_n^2-t^2$, and the states $| x \rangle$ form a distributional basis of the auxiliary Hilbert space (localized delta functions),  with $\langle x|y\rangle = \delta(x-y)$.

We now claim  that  Eq. (\ref{newrigging}) also defines a direct product with the physical, normalized,  inner product. Moreover, we will show that each ${\cal H}_i$ defines a superselected sector of the theory, that is, for any two states $|\phi_i\rangle$, $|\phi_j\rangle$ belonging to different sectors ${\cal H}_i$ and ${\cal H}_j$ respectively, and for any observable $\hat{O}$ in $\cal A$,
\begin{equation}
\langle\phi_i|\hat{O}|\phi_j\rangle_{\mbox{\tiny phy}} =0 \ .
\label{superselection}
\end{equation}
In fact,
\begin{equation}
\lim_{\lambda\rightarrow\infty}\int_{G_\lambda} \langle \phi_i|\hat{O}U(g)|\phi_j \rangle \ dg =
\langle \phi_i|\hat{O}|\phi_j \rangle_{\mbox{\tiny phy}}\Lambda^k(\lambda),
\label{aaa}
\end{equation}
for some $k$. Now let us decompose
\begin{eqnarray}
\hat{O}|\phi_i\rangle &=&  \alpha|\phi_i^{\prime}\rangle + \beta |\phi_j^{\prime}\rangle+|\psi\rangle \\
\hat{O}|\phi_j\rangle &=&  a|\tilde{\phi}_i\rangle + b |\tilde{\phi}_j\rangle+|\tilde{\psi}\rangle \ ,
\label{decomp}
\end{eqnarray}
where the indices $i,j$ indicate where those states live, $(a,b,\alpha,\beta)$ are complex numbers and $|\psi\rangle$,
$|\tilde{\psi}\rangle$ live in the  ortogonal complement of  ${\cal H}_i \oplus{\cal H}_j$.
 As $\hat{O}$ is selfadjoint in
 ${\cal H}_{\mbox{\tiny aux}}$ we have
 \begin{eqnarray*}
\langle\phi_i| \hat{O}|\phi_j\rangle_\lambda &=& \beta^{*}\langle\phi_j^{\prime}|\phi_j\rangle_{\lambda} \\
&=&  a \langle\phi_i|\tilde{\phi_i}\rangle_{\lambda} \ \ .
 \end{eqnarray*}
Therefore we see that the right hand side expressions  must vanish in the $\lambda \rightarrow \infty$ limit. Otherwise,
we will get a contradiction, for by definition they have different degree of divergency.

Finally, note that the inner product defined in (\ref{newrigging}) is not unique if there are superselected sectors.
There is always an undetermined constant factor when one drops out the infinite factors $\Lambda^i$. When only one sector  is present,  this factor is given by the normalization of the states, but otherwise, we can always define different weights
in different sectors, giving us a freedom in the election of the inner product.

\section{Conclusions}

\label{conc}

We have seen that in certain circumstances it is posible to generalize the group averaging method in order
to extend its applicability to non--compact groups.  This is done by regularizing the integrals and dropping out the infinite
factors that might occur. When the auxiliary Hilbert space contains different sectors for which the integrals have different
degrees of divergency,  the physical Hilbert space will be a direct sum of these sectors:
$$
{\cal H}_{\mbox{\tiny phy}}= \bigoplus_{i}{\cal H}_{\mbox{\tiny phy}}^{i} \ \ ,
$$
with the inner product for each sector defined in (\ref{newrigging}). Each ${\cal H}_{\mbox{\tiny phy}}^{i}$ is
superselected (no observable  has matrix element between their elements). Because of this, the inner product is defined inside
each sector independently and we can multiply it by different weights, giving us a  freedom in the definition of the inner product
for the whole physical Hilbert space.

\section*{Acknowledgements}

The author wishes to thank Donald Marolf  for many discussions and
comments.
This work was supported in part by the NSF grant PHY97-22362,  by   Fundaci\'on Andes and
funds from Syracuse University.
The institutional support to CECS provided by  I. Municipalidad de Las Condes and a group
of Chilean companies (AFP Provida, CODELCO, Empresas CMPC,
MASISA S.A. and Telef\'onica del Sur) is also acknowledged.
CECS is a Millennium Science Institute.

%


\begin{thebibliography}{100}
%

\bibitem{ggmm} A.~Gomberoff and D.~Marolf,
Int.\ J.\ Mod.\ Phys.\  {\bf D8}, 519 (1999) ,
[gr-qc/9902069].

\bibitem{Dirac} Dirac, P A M {\it Lectures on Quantum Mechanics}
(New York: Belfer Graduate School of Science, Yeshiva University, 1964).

\bibitem{HT} M. Henneaux and C. Teitelboim {\it Quantization
of Gauge Systems} (Princeton University Press, Princeton, 1992).



\bibitem{ALMMT}
A. Ashtekar, J. Lewandowski, D. Marolf,
J. Mour\~ao, and T. Thiemann,
{\it J.~Math. Phys.} {\bf 36} 6456 (1995), gr-qc/9504018.

\bibitem{GM} D.~Giulini and D.~Marolf,
Class.\ Quant.\ Grav.\  {\bf 16}, 2479 (1999),
[gr-qc/9812024].


\bibitem{KL}
 N. Landsman,
{\it J. Geom. Phys.} {\bf 15} (1995) 285-319, hep-th/9305088.

\bibitem{AH}  A. Higuchi,
{\it Class. Quant. Grav.} {\bf 8}  (1991) 1983.

\bibitem{QORD} D. Marolf,
Class. Quant. Grav. {\bf 12} (1995) 1199,
gr-qc/9404053.

\bibitem{GM2} D.~Giulini and D.~Marolf,
Class.\ Quant.\ Grav.\  {\bf 16} (1999) 2489 ,
[gr-qc/9902045].


\bibitem{BC} D. Marolf, in
{\it Symplectic Singularities and Geometry of Gauge Fields},
(Banach Center Publications, Polish Academy of Sciences, Institute
of Mathematics Warsaw, 1997); gr-qc/9508015.

\end{thebibliography}
\end{document}